\documentclass[12pt]{article}

\begin{document}

\author{E. Ahmed and M. F. Elettreby \\Mathematics Department, Faculty of Science,\\
Mansoura University, Mansoura 35516,
Egypt\\mohfathy@mans.edu.eg\\(to be appear in Int. J. Mod. Phys.
C)}

\title{On Multiobjective Evolution Model}

\maketitle

\begin{abstract}

Self-Organized Criticality (SOC) phenomena could have a
significant effect on the dynamics of ecosystems. The Bak-Sneppen
(BS) model is a simple and robust model of biological evolution
that exhibits punctuated equilibrium behavior. Here we will
introduce random version of BS model. Also we generalize the
single objective BS model to a multiobjective one.
\end{abstract}


\section{Introduction}

In recent years, there has been increasing interest of the
possibility that evolution of species in an ecosystem may be a
self organized critical phenomena. There are interactions among
species in that ecosystem. The most common such interactions are
predation, competition for resources and mutualism. As a result of
these interactions the evolutionary adaptation of one species must
affect its nearest neighbors. These interactions can give rise to
large evolutionary disturbances, termed \textit{coevolutionary
avalanches}. Most of these evolution models, like BS model [1],
considered only one fitness for each species. Bak and Sneppen
proposed a self organized model to explain the punctuated
equilibrium of biological evolution. They considered a 1-
dimensional model with periodic boundary conditions, topologically
a circle. Assign a fitness $0<f\left( i\right)<1$ to each site
$i$, $i=1,2,...,N$, where $N$, is the number of species in the
ecosystem. At each time step look for the site with lowest fitness
$j$ then replace its fitness together with the fitnesses of its
nearest neighbors, $j\pm 1,$ by new ones which are uniformly
distributed random variables.

\begin{eqnarray*}
f\left( j\right) &=&random\,value \\
f\left( j+1\right) &=&random\,value \\
f\left( j-1\right) &=&random\,value
\end{eqnarray*}

After running the system for sufficiently long time most of the
fitness are above certain threshold $.667$. Also, the distribution
of the distance between subsequent mutations and the avalanche
sizes exhibit power laws.

Several modification can be done to the BS model. The first
possible modification is to use extremal dynamics [2, 3] which
depends on the following idea: In real biological systems not only
the lowest one who is updated but some of the low fitness species.
This number that changes is not fixed but random. So, we will
study this random version of BS model in section 2. Also, in
biology almost every optimization problem is multiobjective (MOB)
e.g. objective of foraging and of minimizing predation risk. In
section 3, we will apply the concept of MOB to BS model.

\section{Random BS Model}

Here, we will study the first modification that can be done to the
BS model. Instead of finding exactly the site with lowest fitness,
one may use the extremal dynamics. In this case a uniformly
distributed random number is picked and all the sites with fitness
less than this number has its fitness updated. This dynamics has
been used to explain the long term memory for the immune system
[3]. It has been also used to solve some optimization problems
e.g. spin glass, graph coloring and graph partitioning [4, 5]. We
run a system consisting of $N=4096$ species for different
sufficiently long time (up to $2\times 10^{7}$). We find that most
of the fitnesses are above a certain threshold value $0.64$, as
shown in figure 2. In figure 1 we plotted the standard Bak-Sneppen
model for reference.

\section{Multiobjective Optimization Model}

In most evolution models e.g. BS model, only one fitness is
considered i.e. single objective optimization. Almost every real
life problem is multiobjective (MOB) one [6]. Therefore it is
important to generalize the standard single goal oligopoly studies
to multiobjective ones. Methods for MOB optimization are mostly
intuitive.

The \textbf{first method} is lexicographic method. In this method
objectives are ordered according to their importance. Then the
first objective is satisfied fully. The second one is satisfied as
much as possible given that the first objective has already been
satisfied and so on. A famous application is in university
admittance where students with highest grades are allowed in any
college they choose. The second best group are allowed only the
remaining places and so on. This method is useful but in some
cases it is not applicable.

The \textbf{second method} is the method of weights [7]. Assume
that it is required to minimize the objectives $Z(i)$,
$i=1,2,...,N$. The problem of maximization is obtained via
replacing $Z(i)$ by $-Z(i)$. Define
\[Z=\sum_{i=1}^{N}\,w(i)\,Z(i)\]
where
\[w(i)\geq 0\,\,\,\,\,{\rm and}\,\,\,\,\,\,\sum_{i=1}^{N}w(i)=1\]
Then the problem becomes to minimize $Z$. This method is easy to
implement but it has several weaknesses. The first is that it may
give a Pareto dominated solution. A solution $Z^{^{\prime
}}(i),\,\,\,\,i=1,2,\,...,N$ is Pareto dominated if there is
another solution $Z(i),\,i=1,2,\,...,N$ such that $Z(i)\leq
Z^{^{\prime }}(i)$ for all $i$ with at least one $k$ such that
$Z(k)<Z^{^{\prime }}(k)$. The second difficulty of this method is
that it is difficult to apply for large $N$.

The \textbf{third method} is to minimize only one objective while
setting the other objectives as constraints e.g. minimize $Z(1)$
subject to $Z(i)\leq a(i)$, $i=2,3,...,N$ where $a(i)$ are
parameters to be updated. The problem with this method is the
choice of the thresholds $a(i)$. In the case of equality i.e.
$Z(i)=a(i)$ this method is guaranteed to give a Pareto optimal
solution.

The \textbf{fourth method} using fuzzy logic is to study each
objective individually and find its maximum and minimum say
$Z_{\max }(i)$, $Z_{\min}(i)$ respectively. Then determine a
membership
\[m(i)=\frac{Z(i)-Z_{\max }(i)}{Z_{\max }(i)-Z_{\min }(i)}\]
Thus $0\leq m(i)\leq 1$. Then apply $\max\left(
\min\left(m(i),\,i=1,2,...,N\right) \right)$. Again this method is
guaranteed to give a Pareto optimal solution. This method is a bit
difficult to apply for large number of objectives.

The BS model can be generalized to the multiobjective. Assigning
two fitnesses $f_{1}\left( i\right)$, $f_{2}\left( i\right)$, to
each site instead of one. The updating rule is If
\[x\;\,f_{1}\left( i\right) +\left( 1-x\right) \;\,f_{2}\left( i\right) <\min{\rm fit}\]
where $0<x<1$, then update both $f_{1}\left( i\right)$ ,
$f_{2}\left(i\right)$ and $f_{1}\left( i\pm 1\right)$,
$f_{2}\left( i\pm 1\right)$.

In the updating rule we have used the simple and widely used
method, weighting method in MOB. Multiobjective optimization is
much more realistic than single objective ones. After running a
system consisting of $N=4096$ species for different sufficiently
long time (up to $2\times 10^{7}$). The distribution of the
distance between subsequent mutations are shown in figure 3. We
find that most of the fitnesses are above certain threshold value
$0.57$, as shown in figure 4. The size of avalanches are shown in
figure 5.

\newpage
\begin{center}
\bigskip
\bigskip
FIGURES
\end{center}

\begin{enumerate}

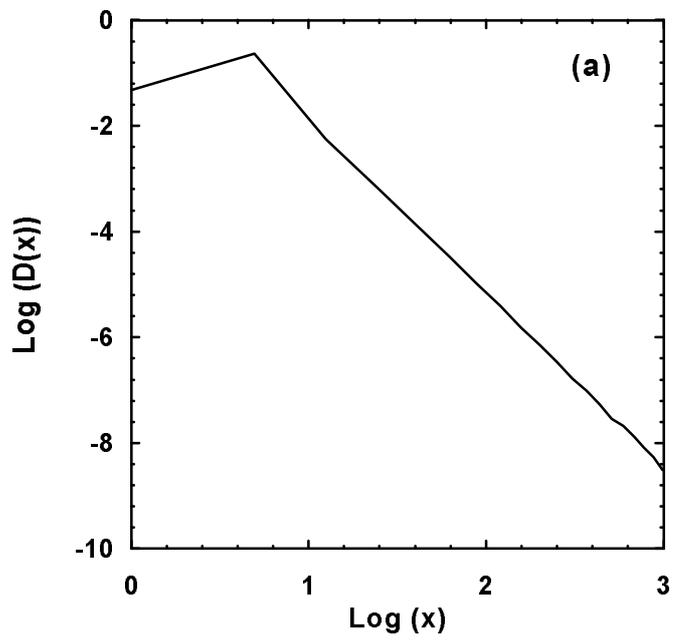
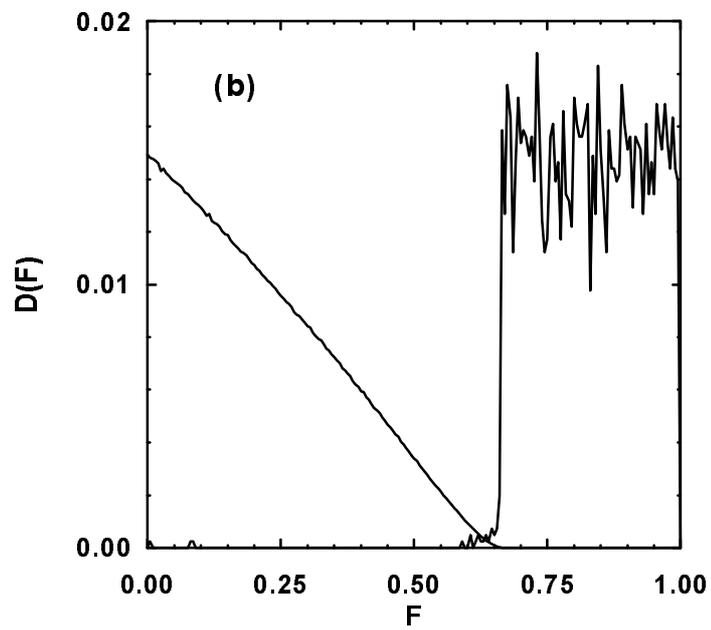
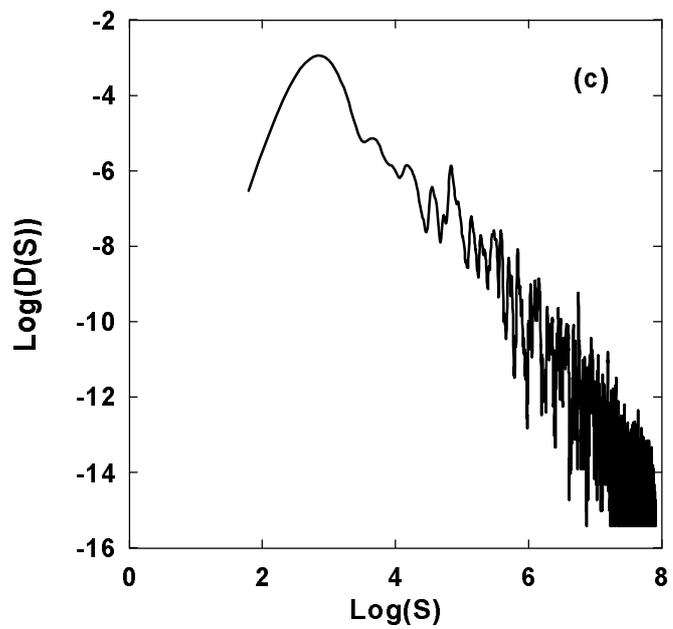
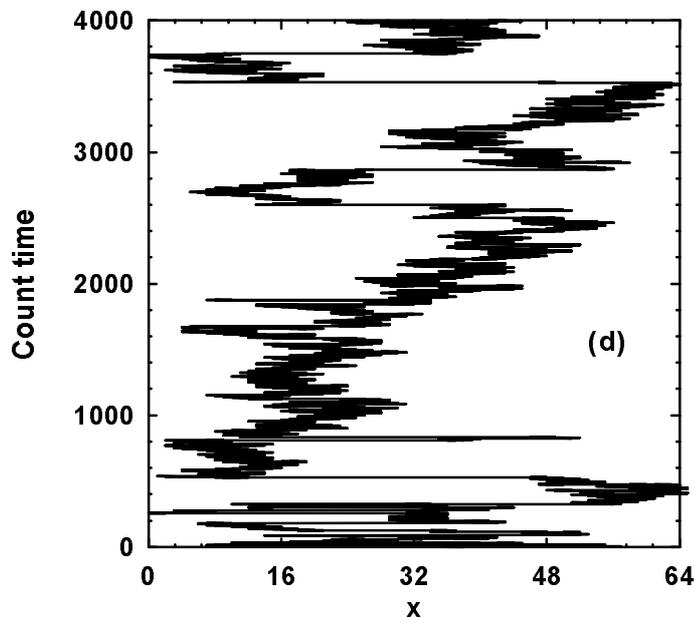
\item[FIG. 1.]  The results of BS model under our simulations in a system of
size $N=4096$ and $t=10^{7}$, iteration. (a) The distribution of
distances D(x) between subsequent mutations x. (b) Distribution of
fitness in the critical state (right curve) D(F) with the
distribution of minimum fitness (left curve). (c) Distribution of
avalanche sizes D(S) in critical state. (d) Mutation activity vs
time measured as the total number of mutations ($N=64$ and
$t=4\times 10^{3}$).
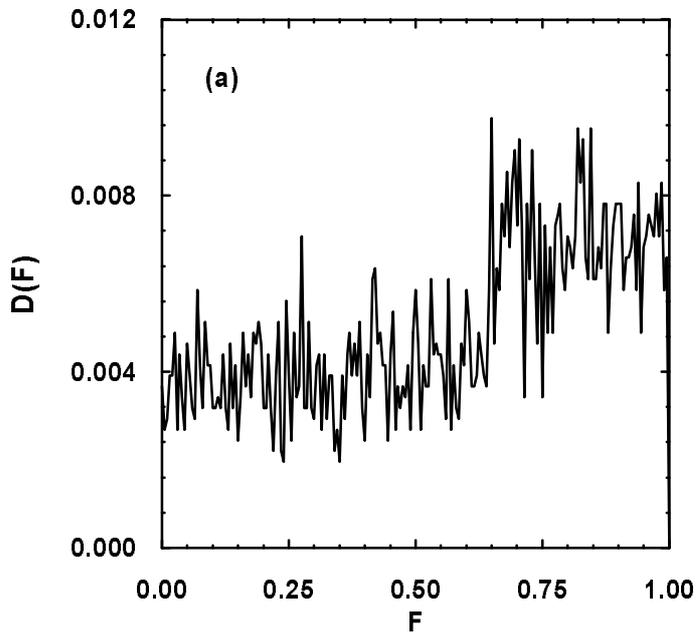
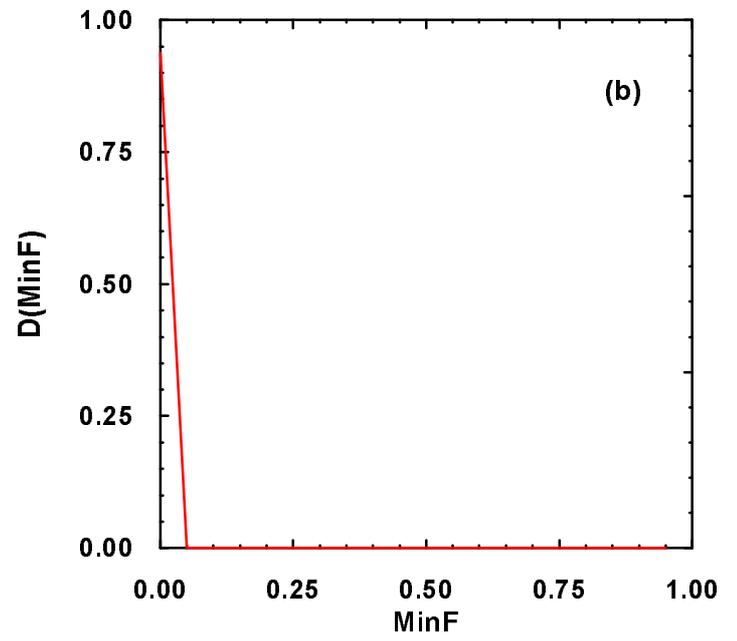
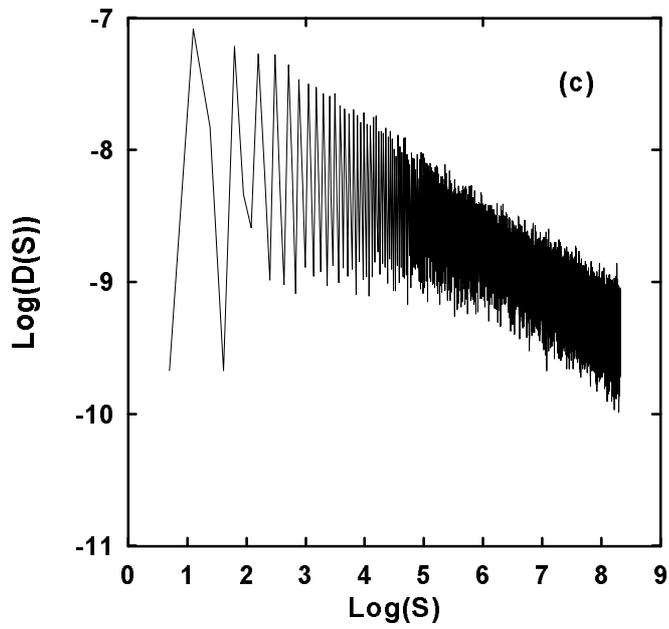
\item[FIG. 2.]  The results of random BS model in a system of size $N=4096$
and $t=10^{7}$, iteration. (a) Distribution of fitness in the
critical state D(F). (b) Distribution of minimum fitness in the
critical state. (c) Distribution of avalanche sizes D(S) in
critical state.
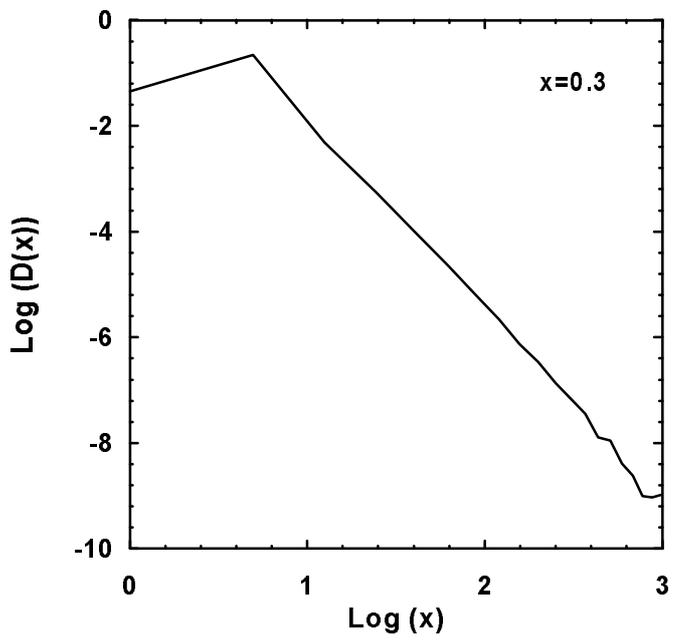
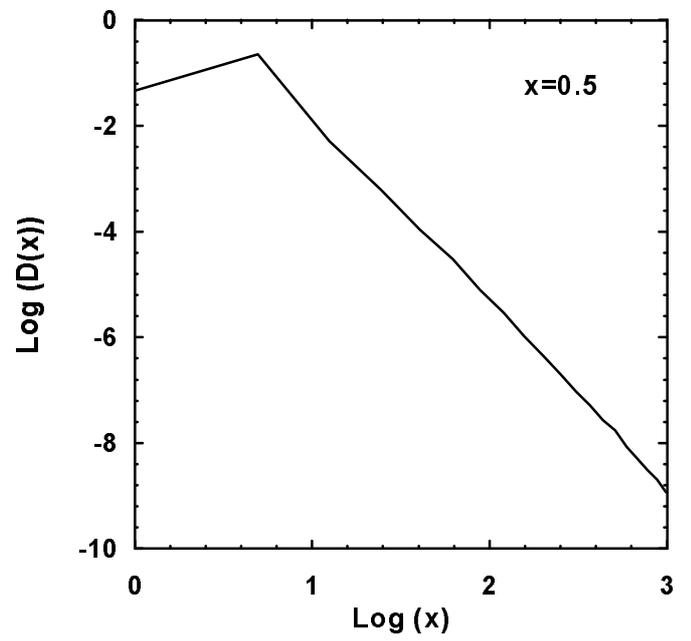
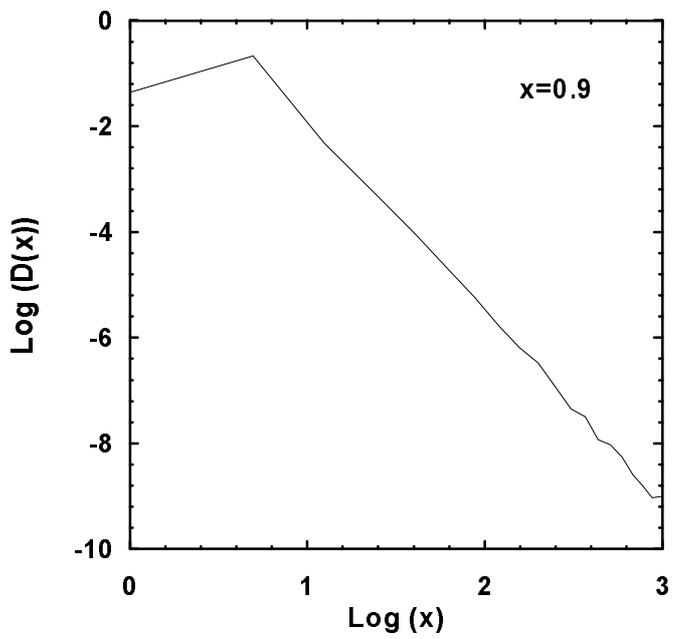
\item[FIG. 3.]  The distribution of distances D(x) between subsequent
mutations in a system of size $N=4096$ and $t=10^{7}$, iteration
with three different weights 0.3, 0.5, 0.9.
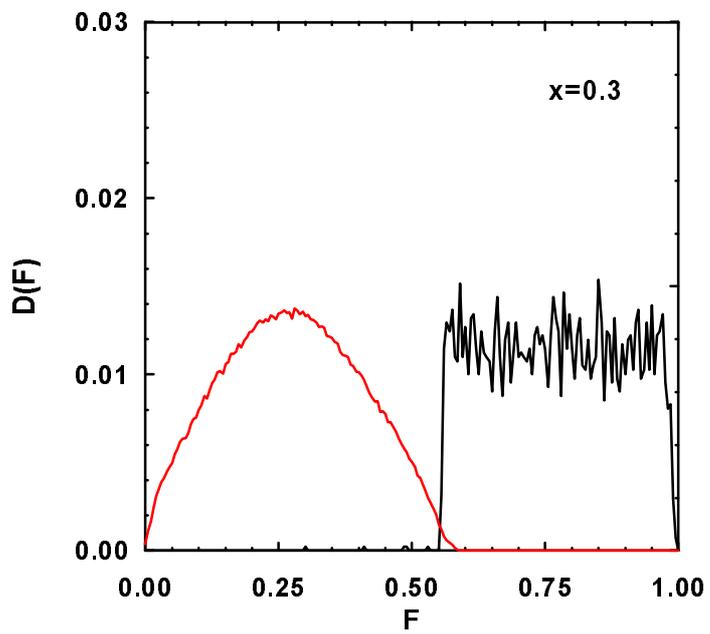
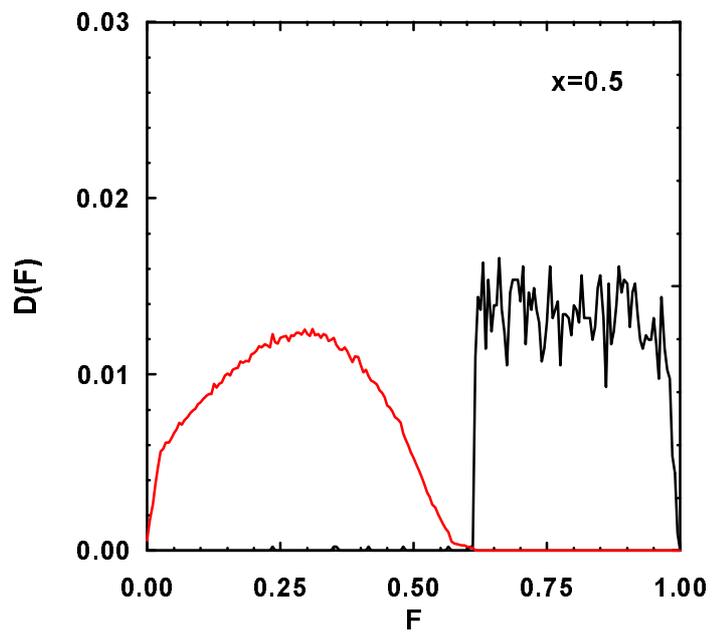
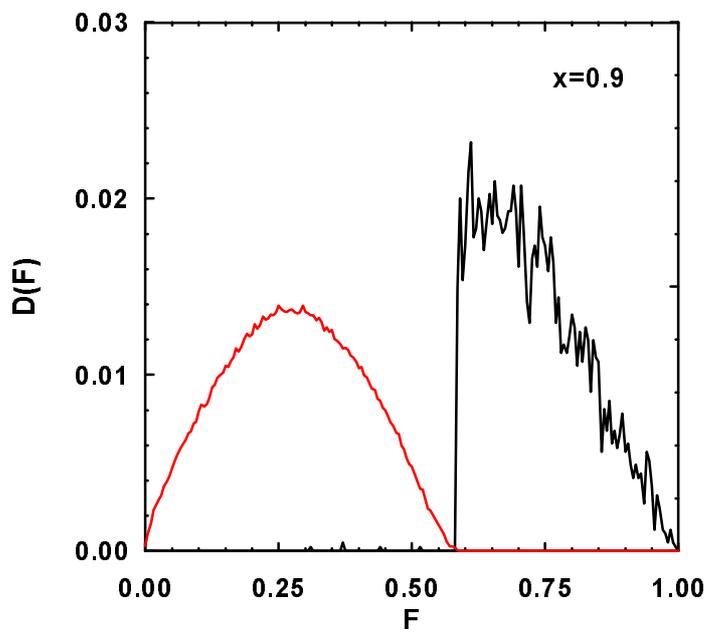
\item[FIG. 4.]  The distribution of fitness in the critical state (right
curve) D(F) with the distribution of minimum fitness (left curve)
in a system of size $N=4096$ and $t=10^{7}$, iteration with three
different weights 0.3, 0.5, 0.9.
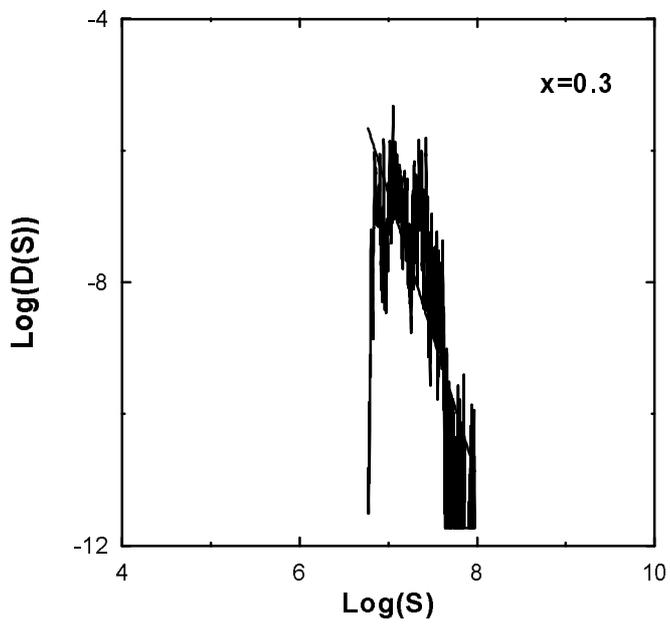
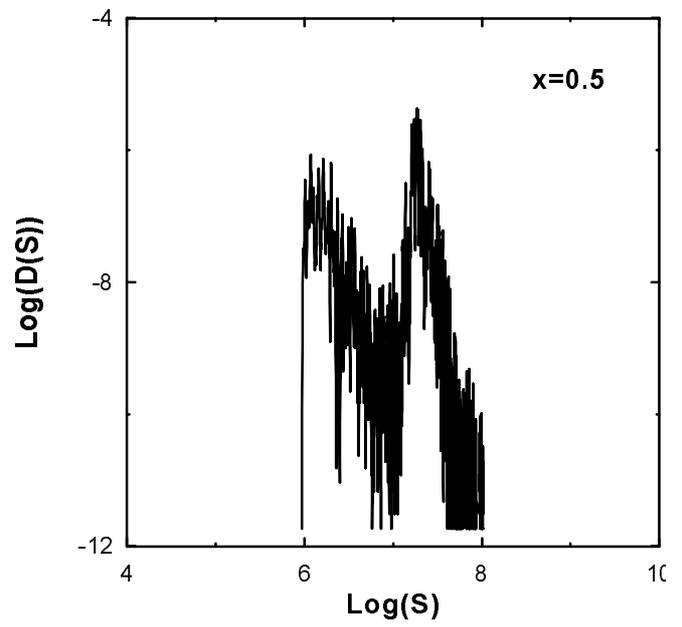
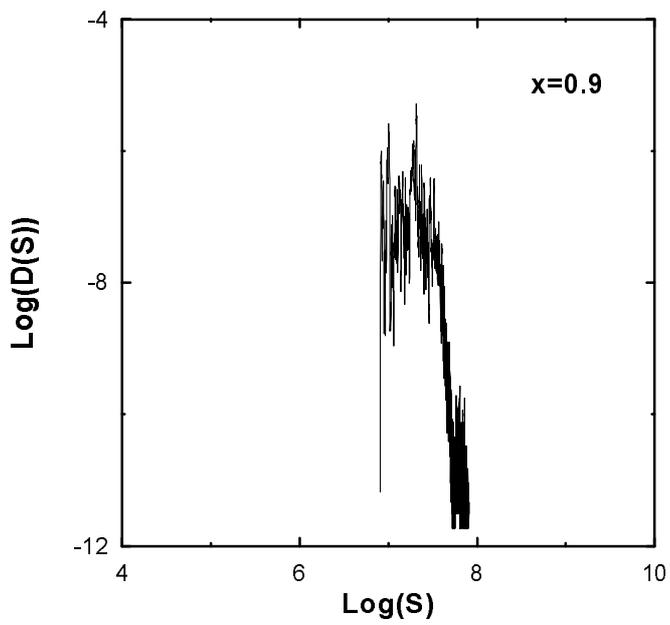
\item[FIG. 5.]  The distribution of avalanche sizes D(S) in critical state
in a system of size $N=4096$ and $t=10^{7}$, iteration with three
different weights 0.3, 0.5, 0.9.

\end{enumerate}

\end{document}